\documentclass[aps,prl,twocolumn,showpacs,groupedaddress,nofootinbib]{revtex4}
\usepackage{graphicx}
\begin{document}

\title{Study of Possible Proton-Antiproton Hexaquark State in Lattice QCD}
\author{Mushtaq Loan}
\affiliation{International School, Jinan University, Huangpu Road
West, Guangzhou 510632, P.R. China}
\date{July 18, 2008}
\begin{abstract}
We have used standard techniques of lattice quantum chromodynamics
to look for  evidence of the spin-zero six quark flavour singlet
state ($J^{PC}=0^{-+}$) observed by BES Collaboration, and to
determine the splitting between the mass of the possible
proton-antiproton and the mass of two protons, its threshold.
Using the various interpolating fields in the quenched
approximation we find indications that for sufficiently light
quarks proton-antiproton is slightly above the $2m_{p}$ threshold.
The lattice resonance signal of binding observed near the physical
and continuum regimes do not support the existence of
proton-antiproton state as a spin-zero hexaquark state in quenched
lattice QCD.

\end{abstract}

\maketitle

Recently, the BES Collaboration in Beijing observed a near
threshold enhancement in the proton-antiproton $(p\bar{p})$ mass
spectrum from  $J/\psi \rightarrow \gamma p\bar{p}$ radiative
decay \cite{Bai2003}. Fitting the enhancement with  an $S$- wave
Breit-Wigner resonance function, results in peak mass at
$M=1859^{+3}_{10}(\mbox{stat})^{+5}_{25}(\mbox{sys})$ with a total
width $\Gamma < 30$ $\mbox{MeV}/\mbox{c}^{2}$. With a $P$-wave
fit, the peak mass is very close to the threshold, $M=1876.4\pm
0.9$ MeV and the total width is very narrow, $\Gamma = 4.6\pm 1.8$
MeV. This discovery is subsequently confirmed  by the Belle
Collaboration in different reactions of the decays
$B^{+}\rightarrow K^{+}p\bar{p}$ \cite{Abe2002} and
$\bar{B}^{0}\rightarrow D^{0}p\bar{p}$ \cite{Abe2002b}, showing
enhancements in the $p\bar{p}$ invariant mass distribution near
$2m_{p}$. Such a mass and width does not match that of any known
particle \cite{pdata2002}.

Theoretical existence of possible proton-antiproton  bound state
has long been  speculated in quark model and conventional nucleon
potentials \cite{Klempt2002,Richard2000,Datta2003,Rosner2003}.
However, it was only in a very recent study, made  by Yan \emph{et
al} using the Skyrme model \cite{Yan2005}, that predicted mass and
the width close with the experiment. Experimentally, the quantum
numbers of $p\bar{p} (1857)$ are not well determined yet. The
photon polar angle distribution was found consistent with
$1+\mbox{cos}^{2}\theta_{\gamma}$ suggesting the angular momentum
is very likely to be $J=0$. Making full use of general symmetry
requirement and available experimental information the
corresponding spin and parity are $J^{PC}=0^{-+}$ \cite{Gao04}. In
this letter we report first quenched lattice QCD calculation
capable of studying the $p\bar{p} (1859)$.

In contrast with conventional baryons and mesons, it is difficult
to deal with $q^{m}\bar{q}^{n}$ ($m+n\geq 3$) states in lattice
QCD. For example, a $q^{3}\bar{q}^{3}$ state can be decomposed
into a couple of colour singlets states even in the absence of
unquenched effect. The two-point function, in general, couples not
only to the single hadron  but also to the two-hadron states. We
seek operators that have a little overlap with the hadronic
two-body states in order to identify the signal of our hexaquark
state in lattice QCD. For this purpose, we first consider a non
two-body-type field for $I(J^{PC})=0(0^{-+})$ and construct our
interpolating fields based on the diquark description.

In Jaffe model \cite{Jaffe05}, each pair of $[ud]$ form a diquark
which transforms like a spin singlet ($1_{s}$), colour
anti-triplet ($\bar{3}_{c}$), and the flavour anti-triplet
($\bar{3}_{f}$). Therefore, for  a diquark operator, one has
\cite{Sasaki04}
\begin{equation}
Q^{i,a}_{\Gamma}(x)=
\frac{1}{2}\epsilon_{ijk}\epsilon_{abc}q^{T}_{j,b}(x)C\Gamma
q_{k,c}(x),
\label{eqn01}
\end{equation}
where $\epsilon_{abc}$ is completely antisymmetric tensor, and
$(a,b,c)$ and $(i,j,k)$ denote the colour and flavour indices,
respectively.  The superscript $T$ denotes the transpose of the
Dirac spinor and $C$ is the charge conjugation matrix. The Dirac
structure of the operator is represented by the matrices $\Gamma$,
satisfying $\Gamma_{\alpha\beta}=-\Gamma_{\beta\alpha}$ ($\alpha$
and $\beta$ are Dirac indices) such that the diquark operator
transforms like a scalar or pseudoscalar. The colour and flavour
antisymmetries restrict the possible $\Gamma$'s within 1,
$\gamma_{5}$ and $\gamma_{5}\gamma_{\mu}$. Then the hexaquark
hadron $p\bar p$($[ud][\bar{u}\bar{d}][\bar{u}u]$) emerges as a
member with $S=0$ and $I=0$ in $(Q^{\bar3}\otimes
\bar{Q}^{3})\otimes Q^{6} = ([\bar{15}_{cs}]\otimes
[15_{cs}])\otimes [21_{cs}]$ in SU(6) colour-spin representation
and a flavour singlet in $(\bar{3}_{f}\otimes 3_{f})\otimes
6_{f}$. With this picture the interpolating operator for $0^{-+}$
is obtained as
\begin{equation}
\chi_{dq}(x) = \epsilon_{abc}Q^{i,a}_{\Gamma}(x){\bar
Q}^{i,b}_{\Gamma^{'}}(x)Q'^{s,c}_{\Gamma''}(x) \label{eqn02a}
\end{equation}
where $Q^{'} = u^{T}C^{-1}\gamma_{5}\Gamma^{''}u$. This
identification looks familiar if we represent one of the quarks by
charge conjugate field: $q_{a}q_{b} \rightarrow
\bar{q}_{Ca}q_{b}$, where
$\bar{q}_{C}=-iq^{T}\sigma^{2}\gamma_{5}$. Then the classification
of diquark bilinears is analogous to that of $q\bar{q}$ bilinears.
We choose $\Gamma =1$ and $\Gamma^{'}=\Gamma^{''}=\gamma_{5}$.
There are many more possibilities of constructing the operator
even in the $I=0$ channel. In principle testing various other
interpolating operators for the best overlap with $0^{-+}$ state
should provide information on the wave function of the particle.
Our second operator, which we will refer to as the $N\bar
N$-interpolating field, is generalised to obtain an isospin $I=0$
colour-singlet $N\bar N$-type hexa-quark interpolating field:
\begin{equation}
\chi_{N\bar N} = \frac{1}{\sqrt{2}}\left[O_{N}O_{\bar N}-O_{\bar
N}O_{N}\right], \label{eqn02b}
\end{equation}
where
\begin{equation}
O_{N}(x)=\sum_{x}\epsilon_{abc}\left[(u_{a}^{T}(x)C\gamma_{5}d_{b}(x))u_{c}(x)-(u\leftrightarrow
d)u_{c}(x)\right] \label{eqn03}
\end{equation}
and
\begin{equation}
O_{\bar N}(x)= \sum_{x}\epsilon_{abc}\left[({\bar
u}_{a}(x)C\gamma_{5}{\bar{d}}^{T}_{b}(x)){\bar u}_{c}(x)-({\bar
u}\leftrightarrow {\bar d}){\bar u}_{c}(x)\right] \label{eqn04}
\end{equation}
are the standard interpolating fields for the upper two Dirac
components of the nucleon and antinucleon operators, respectively.
These definitions take account of the symmetries for isospin and
nonrelativistic spin representations. In absence of quark
annihilation diagrams,  correlation function is expressed in terms
of basis determined by direct (36 members) and cross (324 members)
contributions of quark contractions. From the dispersion of the
two-point function, the contribution from direct and the cross
amplitudes can be easily calculated. We have taken single
quark-antiquark exchange and a double quark-antiquark exchange to
be equivalent because the $N$ and $\bar N$ operators are summed
over all spatial sites for $s$-wave state.  A more accurate
procedure will be to measure the $2\times 2$ correlation matrix of
two different interpolating operators in Eqs. (\ref{eqn02a}) and
(\ref{eqn02b}) and extract the effective mass from the eigenvalues
using variational techniques \cite{lasscock05}. In this study,
however, we do not intend to pursue this issue further.

To examine the possible proton-antiproton ground state in lattice
QCD, we generate quenched configurations on a $20^{3} \times 60$
lattice (with periodic boundary conditions in all directions) with
tadpole-improved anisotropic gluon action \cite{Morningstar99} in
the coupling range of $3.0 -4.0$. Gauge configurations are
generated by a 5-hit pseudo heat bath update supplemented by four
over-relaxation steps. These configurations are then fixed to the
Coulomb gauge at every 500 sweeps. After discarding the initial
sweeps, a total of 500 configurations for each coupling are
accumulated for measurements. Using the tadpole-improved clover
quark action on the anisotropic lattice \cite{Okamoto02} over the
lattice spacing range of $0.26 - 0.39$ fm, we compute the
light-quark propagators at six values of the hopping parameter
$\kappa_{t}$ which cover the quark mass region of
$m_{u}<m_{q}<2m_{u}$.

To obtain a better overlap with the ground state we used iterative
smearing of gauge links and the application of fuzzing technique
for the fermion fields \cite{UKQCD95}. The application of fuzzing
for two of the six quarks inside the hexaquark state flattens the
curvature of the effective mass. The largest plateau in the region
with small errors is obtained with fuzzed $u$- and $d$- quarks. We
used this variant to calculate our correlation functions.

From the correlation functions we extract the mass (energy) by
standard $\chi^{2}$ fitting with multi-hyperbolic cosine ansatz
\begin{displaymath}
C(t)=\sum_{i=1}^{n}A_{i}\mbox{cosh}(tm_{i}).
\end{displaymath}
However, to ensure the validity of our results, we compared them
with those obtained as the solution to the equation
\begin{displaymath}
\frac{C(t+1)}{C(t)}= \frac{\cosh
[(t+1-N_{t}/2)a_{t}m_{eff}]}{\cosh [(t-N_{t}/2)a_{t}m_{eff}]},
\end{displaymath}
for a given $C(t+1)/C(t)$ at fixed $t$.

\begin{figure}[!h]
\scalebox{0.45}{\includegraphics{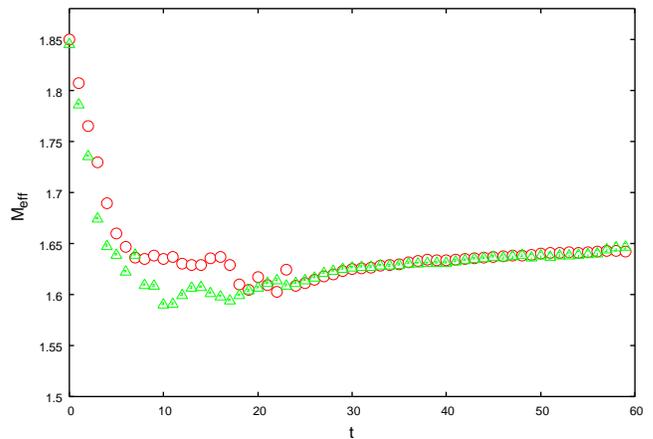}}
\caption{\label{figeffmass} Effective mass of the $J^{PC}=0^{-+}$
colour-singlet state as a function of $t$. The data correspond to
interpolators $\chi_{dq}$ (open circles) and $\chi_{N\bar N}$
(open triangles) at $\beta= 3.5$ and $\kappa_{t}=0.2380$.}
\end{figure}

The fitting range $[t_{min},t_{max}]$ for the final analysis is
determined by fixing $t_{max}$ and finding a range of $t_{min}$
where the ground state mass is stable against $t_{min}$. We choose
one ``best fit" which is insensitive to the fit range, has high
confidence level and reasonable statistical errors. Typical
example of the effective mass plot is shown in Fig.
\ref{figeffmass}. The effective mass decreases monotonically,
which implies that excited spectral contributions are greatly
reduced. For the $\chi_{N\bar N}$-type field, an impressive
plateau with reasonable statistical errors is observed in the
interval $35\leq t\leq 50$, where we expect to achieve single
state dominance.  For the $\chi_{dq}$-type field, the signal is
noisy at earlier times, and hence we fit in the interval $45\leq
t\leq 55$. The best-fit curve to the $M_{N{\bar N}}$ data has
$\chi^{2}/N_{DF} = 1.01$ and for $M_{dq}$, we find
$\chi^{2}/N_{DF}=0.89$. Statistical errors of masses are estimated
by a single elimination jackknife method. We kept statistical
errors under control by ensuring that analyzed configuration are
uncorrelated, which is made possible by separating them by as many
as 500 sweeps. The statistical uncertainties on our hadron masses
are typically on the few percent level. The masses of the
non-strange mesons $\pi$, $\rho$ as well as the nucleon were
computed for scale setting and analyzing the stability of
hexa-quark state, respectively.

\begin{figure}[!h]
\scalebox{0.45}{\includegraphics{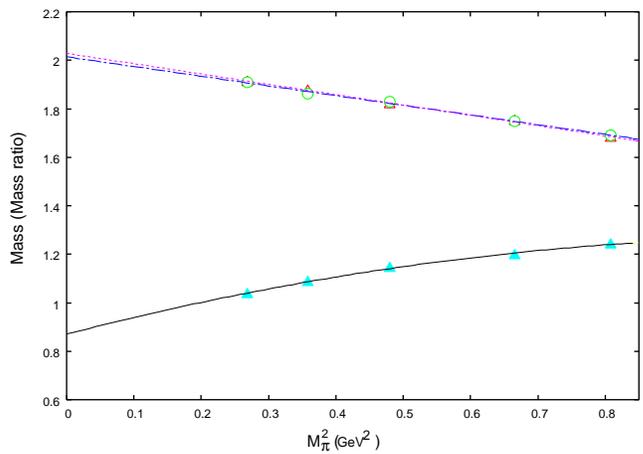}} \caption{
\label{figchiralexp} Chiral extrapolation of hadron  mass ratios,
$M_{N{\bar N},dq}/M_{P}$, extracted with the diquark-type (open
triangles) and $N\bar N$ (open circles) interpolating fields at
$a_{s}=0.32$ fm. The solid triangles correspond to the data from
proton interpolating field. The dashed lines are chiral
extrapolation (linear in $m^{2}_{\pi}$) to the mass ratios using
Eq. (\ref{eqn05}).}
\end{figure}

The chiral extrapolation to the physical limit is the next
important issue. From the view point of chiral perturbation
theory, data points with smallest $m_{\pi}^{2}$ should be used to
capture the chiral log behaviour. Leinweber \emph{et al}
\cite{Derek04} demonstrated that the chiral extrapolation method
based upon finite-range regulator leads to extremely accurate
value for the mass of the physical nucleon with systematic errors
of less than one percent . Unfortunately there does not seem to
exist such a chiral extrapolation technique for multi-quark
hadrons beyond a naive one. Since quenched spectroscopy is quite
reliable for mass ratio of stable particles, it is physically even
more motivated to extrapolate mass ratio instead of mass. This
allows for the cancellation of systematic errors since the hadron
states are generated from the same gauge field configurations and
hence systematic errors are strongly correlated. Fig.
\ref{figchiralexp} collects and displays the resulting particle
mass ratios, $M_{N{\bar N}, dq}/M_{p}$, extrapolated to the
physical quark mass value using linear and quadratic fits in
$m_{\pi}^{2}$
\begin{equation}
m_{h}=a+bm_{\pi}^{2},\hspace{0.10cm}
m_{h}=a+bm_{\pi}^{2}+cm_{\pi}^{4}. \label{eqn05}
\end{equation}
For the nucleon we have employed the procedure adopted in Ref.
\cite{Derek04}. The difference between these two extrapolations
gives some information about systematic uncertainties in the
extrapolated quantities. The data for the mass ratios $M_{N{\bar
N},dq}/M_{p}$ behave almost linearly in $m_{\pi}^{2}$ and both the
linear and quadratic fits, in Eq. (\ref{eqn05}), essentially gave
the identical results. The mass ratios extracted from the fields
$\chi_{dq}$ and $\chi_{N\bar N}$ are very close and agree within
the errors. It can be seen that mass ratios show a weaker
dependence on the quark mass. We expect the contributions from the
uncertainties due to chiral logarithms in the physical limit
significantly less dominant at our present statistics.

\begin{figure}[!h]
\scalebox{0.45}{\includegraphics{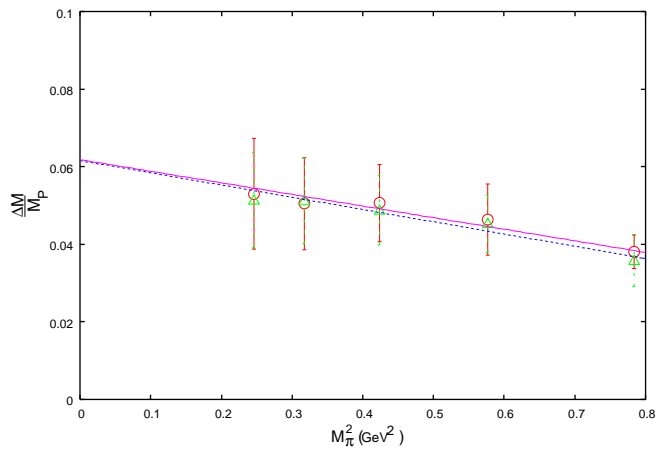}}
\caption{\label{figmassdiff1}Chiral extrapolation of the ratios,
$\Delta M/M_{P} = (M_{N{\bar N},dq}-2M_{P})/M_{P}$, extracted from
diquark-type (open circles) and $N\bar N$ (open triangles) fields
at $a_{s}=0.29$ fm. The solid and the dashed curves are chiral
extrapolation (linear in $m_{\pi^{2}}$) using smallest five
masses.}
\end{figure}

The question whether the lowest-lying $J^{PC}=0^{-+}$ state
extracted is scattering state or a resonance is better resolved by
analysing the ratio of the mass differences $\Delta M = M_{N{\bar
N},dq}-2M_{P}$ (between the candidate low-lying hexa-quark state
and the free two-particle state) and the nucleon mass. The fitted
data, illustrated in Fig. \ref{figmassdiff1}, show that both the
field operators give the consistent estimates of the ratio and the
masses derived from both the operators are consistently higher
than the lowest-mass two-particle state. The ratios are clearly
positive and increase in magnitude  as we approach the physical
regime.
At zero quark mass the mass difference for both the field
operators is $\sim 55(2)$ MeV with $\chi^{2}/N_{DF} = 1.05$.  In
other words, binding becomes weaker near the physical regime with
a general trend of positive binding as the zero quark mass limit
is approached. The chiral uncertainties in the physical limit are
less than $2\%$.

\begin{figure}[!h]
\scalebox{0.45}{\includegraphics{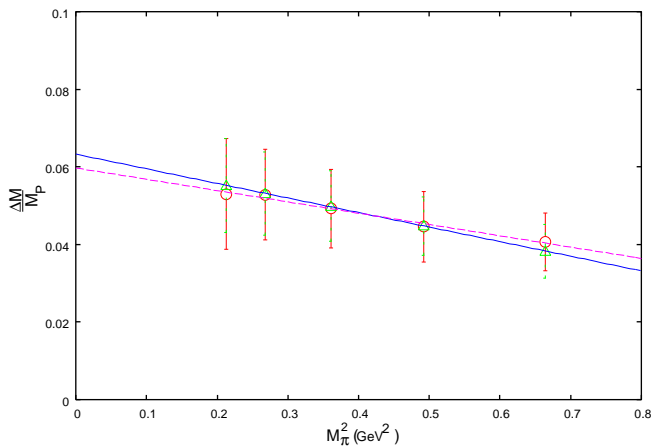}}
\caption{\label{figmassdiff2} As in Fig. \ref{figmassdiff1} but at
$a_{s} =0.26$ fm.}
\end{figure}

A similar behaviour observed for the mass differences between the
hexaquark and the two-particle states at smaller lattice spacing
$a_{s}= 0.26$ fm is illustrated in Fig. \ref{figmassdiff2}. The
ratio obtained from the $\chi_{dq}$ operator for the five smallest
quark masses is slightly higher than that obtained from the
$\chi_{N\bar N}$ operator. For  $N\bar N$-type field the mass
difference is $\sim 56(1)$ MeV  and  $\sim 59(2)$ MeV for
diquark-type interpolator.  The positive mass splitting in both
these channels could be interpreted as the evidence of repulsion.
The chiral uncertainties are again estimated to be less than a
percent.
Similarly, positive mass differences that approximately remained
constant with $m_{\pi}^{2}$ and consistent with those obtained at
our regular lattices was observed on a larger lattice volume of
$6.24^{3}\mbox{fm}^{3}$. Again the signature of repulsion at quark
masses near the physical regime would imply no evidence of the
resonance in the $J^{PC}=0^{-+}$ channel.

\begin{figure}[!h]
\scalebox{0.45}{\includegraphics{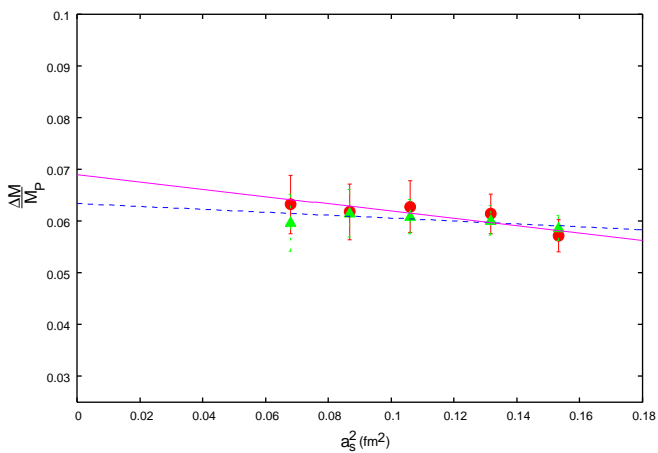}} \caption{
\label{figcontextra} Continuum extrapolation of the mass ratios
extracted from diquark-type (solid circles) and $N\bar N$ (solid
triangles) fields. The solid and the dashed curves are linear
extrapolation to the continuum limit.}
\end{figure}

Finally, we performed a continuum extrapolation for the chirally
extrapolated quantities in Fig. \ref{figcontextra}. Expecting that
dominant part of scaling violation errors  is largely eliminated
by tadpole improvement, we adopt an $a_{s}^{2}$- linear
extrapolation for the continuum limit. We also perform an
$a_{s}$-linear extrapolation to estimate systematic errors.
Performing such extrapolations, we adopt the choice which shows
the smoothest scaling behavior for the final values, and use
others to estimate the systematic errors. As can be seen from Fig.
\ref{figcontextra}, the mass ratio again shows a weak dependence
on the lattice spacing and varies only slightly over the fitting
range. The four non-zero lattice spacing values of the ratio are
within 0.01 - 0.02 standard deviations of the extrapolated zero
lattice spacing result. This will make for unambiguous and
accurate continuum extrapolation.
Using the physical nucleon mass $m_{p}=938$ MeV, we obtain a
continuum mass estimate of $1935(19)(60)$ and $1940(26)(72)$ MeV
for the ground state hexa quark state from $\chi_{N{\bar N}}$ and
$\chi_{dq}$ interpolators, respectively. The results from two
extrapolations seem to be in good agreement, within errors. Given
the fact that the ratio does not show any scaling violations, we
could also quote the value of this quantity on our finest lattice,
which has the smallest error. Nevertheless, order $4\%$ errors on
the finally quoted values are mostly due to the chiral and the
continuum extrapolations. The quenching errors might be the
largest source of uncertainty. Note however, that in the case of
stable hadrons, this is not expected to be very important. It has
been shown \cite{Gattringer03} that with an appropriate definition
of scale, the mass ratios of stable hadrons are described
correctly by the quenched approximation on the $ 1 - 2\%$ level.
To this end we also calculated the pseudoscalar to vector meson
ratio $R_{SP}$ and pseudoscalar to nucleon mass ratio $R_{S N}$
and found that in the continuum these ratios differ about $3\%$
from their corresponding experimental values. So we quote our
quenching errors to be less than five percent. This implies that
dynamical quarks might play a less important role on the spectrum
in question.

Since the mass difference between the reported experimental $p\bar
p$ mass and the physical $2m_{p}$ continuum is $\sim 30$ MeV,
naively one may be tempted to interpret the results in Figs.
\ref{figmassdiff1} and \ref{figmassdiff2} as a signature of the
proton-antiproton bound state on the lattice. But the positive
mass difference observed in the range of pion mass differs from
what is seen for the $p\bar p$ state studied on the lattice. The
continuum results imply that the mass difference does not move
into the continuum with an attractive interaction between $p$ and
$\bar{p}$. This suggests that the observed signal is too heavy to
be identified with the empirical $p{\bar p}(1859)$ and unlikely to
be  a signature of a possible proton-antiproton system as a six
quark state. However the interaction between a baryon and an
antibaryon can be as strong as that of baryon-baryon systems.
Therefore we do not discard the interesting possibility that it
could be a mixed molecular state of baryon and antibaryon bound by
some long-range force of strong or electromagnetic nature or both.

We have presented the results of the first lattice investigation
on the $p\bar p$ state employing improved gauge and fermion
anisotropic actions, relatively light quark masses as well as
smearing  techniques to enhance the overlap with the ground state
of the particle. Our analysis takes into account all possible
uncertainties, such as statistical, finite-size, and quenching
errors when performing the chiral and continuum extrapolations. On
the basis of our lattice calculation  we speculate that the state
is not to be identified as a bound state of six quarks. However a
thorough examination of this question would require the
implementation of flavour SU(3) violation. The $I=0$, $p\bar p$
state couples to $4\pi\eta$ \cite{Gao04} through the $s\bar s$
component of the $\eta$ in the quenched approximation. By giving
the strange quark a larger mass would alter threshold which in
turn would affect the manifestation of the bound state. As
differences between meson masses in full and quenched QCD of order
$60$ MeV \cite{PCAC} and $100$ MeV or more for baryons have been
observed \cite{Young02}, one might wonder what effect the dynamics
of full QCD could have on this state.

Whereas the quenched approximation is found to give quite useful
results at zero temperature, the fermions cannot be neglected if
precise contact with nature is needed. One cannot yet rule out the
possible existence of a $p{\bar p}$ bound state in full QCD.

\begin{acknowledgments}
I would like to thank R. Jaffe, C. Michael, C. Detar, S. Sasaki,
C. Hamer and D. Leinweber for valuable suggestions. It is a
pleasure to acknowledge Shan Jin for his hospitality at CCAST and
stimulating discussions. We are grateful for the access to the
computing facility at the Shenzhen University on 128 nodes of
Deepsuper-21C. The computations were done using a modified version
of the publicly available MILC code (see
www.physics.indiana.edu/sg/milc.html). This work was supported by
the Guangdong Provincial Ministry of Education.
\end{acknowledgments}

\end{document}